\documentclass[aps,prb,showpacs,superscriptaddress,floatfix,twocolumn]{revtex4-1}
\usepackage{times}
\usepackage{graphicx}
\usepackage{amsmath}
\usepackage{amssymb}
\usepackage{subfigure}
\usepackage[colorlinks,citecolor=blue,linkcolor=red]{hyperref}
\usepackage{color}
\usepackage{epstopdf}

\begin{document}

\title{Electronic Correlation and Magnetism in the Ferromagnetic Metal Fe$_3$GeTe$_2$}

\author{Jian-Xin Zhu}
\email[Corresponding author: ]{jxzhu@lanl.gov}
\homepage{http://cint.lanl.gov}
\affiliation{Los Alamos National Laboratory, Los Alamos, New Mexico 87545, USA}

\author{Marc Janoschek}
\affiliation{Los Alamos National Laboratory, Los Alamos, New Mexico 87545, USA}

\author{D. S. Chaves}
\affiliation{Brazil Synchrotron Light Laboratory (LNLS), National Center for Research in Energy and Materials (CNPEM), Campinas, Brazil}
\affiliation{Group of Micro and Nanomagnetism, Institut N\'{e}el, Grenoble, France}

\author{J. C. Cezar}
\affiliation{Brazil Synchrotron Light Laboratory (LNLS), National Center for Research in Energy and Materials (CNPEM), Campinas, Brazil}

\author{Tomasz  Durakiewicz}
\affiliation{Los Alamos National Laboratory, Los Alamos, New Mexico 87545, USA}

\author{Filip  Ronning}
\affiliation{Los Alamos National Laboratory, Los Alamos, New Mexico 87545, USA}

\author{Yasmine Sassa}
\affiliation{Uppsala University, Department of  Physics \& Astronomy, S-75121 Uppsala, Sweden}

\author{Martin Mansson}
\affiliation{Laboratory for Neutron Scattering and Imaging, Paul Scherrer Institute,
CH-5232 Villigen PSI, Switzerland}
\affiliation{Department of Materials and Nanophysics, KTH Royal Institute of
Technology, SE-16440 Stockholm Kista, Sweden}

\author{B. L. Scott}
\affiliation{Los Alamos National Laboratory, Los Alamos, New Mexico 87545, USA}

\author{N. Wakeham}
\affiliation{Los Alamos National Laboratory, Los Alamos, New Mexico 87545, USA}

\author{Eric D. Bauer}
\affiliation{Los Alamos National Laboratory, Los Alamos, New Mexico 87545, USA}

\author{J.~D.~Thompson}
\email[Corresponding author: ]{jdt@lanl.gov}
\affiliation{Los Alamos National Laboratory, Los Alamos, New Mexico 87545, USA}

\begin{abstract}
Motivated by the search for design principles of rare-earth-free strong magnets, we present a study of electronic structure and magnetic properties of the ferromagnetic metal Fe$_3$GeTe$_2$ within local density approximation (LDA) of the density functional theory, and its combination with dynamical mean-field theory (DMFT). For comparison to these calculations, we have measured  magnetic and  thermodynamic properties as well as X-ray magnetic circular dichroism and  the photoemission spectrum of  single crystal Fe$_3$GeTe$_2$. We find that the experimentally determined  Sommerfeld coefficient is enhanced by an order of magnitude  with respect to the LDA value. This enhancement can be partially explained by LDA+DMFT. In addition, the 
 inclusion of dynamical electronic correlation effects provides the experimentally observed magnetic moments, and the spectral density is in better agreement with photoemission data. These results establish the importance of electronic correlations in this ferromagnet.

\end{abstract}
\maketitle

\section{Introduction} 
Electronic correlation effects play an important role in many open $d$- or $f$-shell electronic materials. The competition between kinetic energy and the Coulomb interaction among electrons leads to many emergent phenomena including magnetism and unconventional superconductivity. The emergent functionality has huge potential for technological applications. Magnets have been widely used in different types of devices and motors,~\cite{Lewis12} while superconductors have applications in energy transmission,  high-resolution detectors, and many other technologies.~\cite{LRLawrence:2000} For the former, the magnetocrystalline anisotropy (MA) is one of the most important properties of permanent magnets~\cite{Kirchmayr79} and is large in currently used strong magnets based on rare-earth transition-metal intermetallic compounds, such as  SmCo$_5$ and Nd$_2$Fe$_{14}$B. Recently,  the shortage of rare-earth elements has stimulated the search for rare-earth-free magnetic materials that could be technologically useful  by harnessing sources of magnetic anisotropy other than that provided by the rare-earth components.~\cite{Lewis12} The $3d$, $4d$ and $5d$ transition metals are natural candidates for the search. There the spin-orbit coupling is central for the generation of the orbital component of the magnetic moment,~\cite{JHvanVleck:1937} which together with the lattice anisotropy, determines the contributions of itinerant ferromagnetic electrons to the MA.  

In the past, extensive band structure calculations have been carried out to estimate the magnetocrystalline anisotropy energy (MAE) in ferromagnets. Although those earlier band structure calculations have given the  correct order of magnitude of MAE for certain transition metal ferrogmagnets,~\cite{Bloch31,Brooks40} the calculated MAE usually does not agree quantitatively with experiment~\cite{Daalderop90,JTrygg:1995} and, even more significantly, an incorrect easy axis has been predicted for some systems. This failure was either ascribed to  the implementation challenge encountered in the early stage of electronic structure methods to calculate reliably an energy difference as small as  0.1~meV~\cite{Daalderop90}  or to the inadequacy of the exchange-correlation functional central to density functional theory (DFT).~\cite{Nordstrom92,Jansen90} With the significant progress made in the intervening years, the implementation of DFT-based electronic structure approaches has improved substantially, and this has allowed more attention to be focused on the electronic structure theory itself. In this aspect, YCo$_5$ has been a good case study, providing a rare-earth-free ferromagnet with a MAE comparable to  SmCo$_5$. It has an easy-axis along the $c$-axis of its hexagonal lattice structure, which contains two nonequivalent Co sites. Earlier band-structure calculations~\cite{Nordstrom92} have shown that the orbital moment obtained from LDA  is underestimated by a factor of two when compared with experiments but the inclusion of  an orbital-polarization (OP) potential improves the comparison. Fundamentally, the OP effect is a consequence of the Coulomb interaction between the open-shell $d$- (or $f$-) electrons on the same ion, which modifies the orbital moment and the MAE.  However, the on-site Coulomb interaction should also renormalize electronic band states and this effect goes beyond the description of LDA with OP. Very recently, we have combined  LDA with dynamical mean-field theory (DMFT) to study the magnetic properties in YCo$_5$ and have demonstrated that the incorporation of the electronic correlations  leads to a reliable estimate of the orbital moment, as well as good approximations of both the mass enhancement and the MAE.~\cite{JXZhu:2014}  That study strongly suggested that the electronic correlation effects should be considered as an important part of design principles for itinerant rare-earth-free strong magnets.

This notion stimulated us further to study the role of  strong electronic correlations  in other 3$d$-electron itinerant ferromagnets with similar crystal structure. The ternary compound Fe$_3$GeTe$_2$ has been reported  to be an itinerant ferromagnet,~\cite{NKhAbrikosov:1985,HJDeiseroth:2006} and we have found that it is an easy-axis ferromagnet with an MAE at 5 K that is about 20\% of that of YCo$_5$.~\cite{JXZhu:2014}  In addition, Fe$_3$GeTe$_2$ is particularly interesting in that its crystal structure is built from  Fe$_3$Ge heterometallic slabs that are sandwiched between two Te layers, the latter of which alternate along the $c$-axis of a hexagonal unit cell. Therefore, Fe$_3$GeTe$_2$ provides an opportunity to study the magnetic and electronic properties in this quasi-two-dimensional (2D) itinerant ferromagnet. Its study is  also  interesting in the context of other quasi-2D Fe-based materials, which have been found to be antiferromagnetic bad metals and become superconductors upon chemical doping.~\cite{YJUemura:2009,JDai:2009,PJHirschfeld:2011}  Here we report  a theoretical and experimental study of Fe$_3$GeTe$_2$  that includes  LDA+DMFT electronic structure calculations, as well as  magnetic,  thermodynamic, X-ray magnetic circular dichroism (XMCD), and photoemission spectroscopy measurements.  Our results indicate that Fe$_3$GeTe$_2$ is a strongly correlated ferromagnetic metal and that quantum fluctuation effects are crucial for a correct description of this compound.

The remainder of this paper is organized as follows: In
Sec.~\ref{sec:model} we give details of the theoretical and experimental methods. 
The correlations effects are discussed by comparing the theoretical results of density of states and magnetic moments  with thermodynamic, XMCD, and photoemission spectroscopy measurements in Sec.~\ref{sec:results}. Finally, a brief summary is given in
Sec.~\ref{sec:summary}.

\section{Theoretical and experimental methods}  
\label{sec:model}
Single crystals were grown via iodine vapor transport as previously described.~\cite{BChen:2013}  
Refinement of single-crystal X-ray diffraction at room temperature confirmed  the expected hexagonal structure type $P6_3/mmc$, which contains two inequivalent Fe sites Fe1 and Fe2 and is illustrated in the inset of Fig.~\ref{fig:COT}. Previous studies have indicated deficiency of Fe on the Fe2 site.~\cite{AFMay:2016}  Our refinement also finds an Fe2 deficient occupancy of 0.866 but full occupancy of Fe1 as well as  Ge and Te sites. Lattice parameters for our crystal are $a=b=4.0042(15)$ \AA\;  and $c=16.282(6)$ \AA. Throughout the work,  we will refer to Fe$_{2.87}$GeTe$_2$ as simply Fe$_3$GeTe$_2$.

The specific heat  and magnetic properties were measured on these crystals. 
Synchrotron-based PES measurements of the electronic structure were performed at the Swiss Light Source (SLS) on beamline SIS-X09LA. The energy resolution was set to be better than 15 meV and samples were cleaved in situ and measured at 15 K in a vacuum better than $10^{-10}$ Torr. The surfaces were very stable and without signs of degradation over a typical measurement period of 20 hours.
The XMCD measurements were carried out in a total electron yield detection scheme at the beam line PGM at the Laboratorio Nacional de Luz Synchrotron (Brazilian Synchrotron Light Laboratory).~\cite{JCezar:2013} 
The sample was post cleaved under $10^{-8}$  mbar and measured
under  $\sim 5\times 10^{-9}$  mbar vacuum conditions.   All measurements were carried out at
magnetic field of 20 kOe and temperate $T = 45$ K. The scans were made over an energy range of 695 to 740 eV to measure the Fe $L_3$ and $L_2$ edges (706.8 and 719.9 eV, respectively).

The experimental crystal structure parameters of Fe$_3$GeTe$_2$
were used for the electronic structure and magnetic properties calculations.  
The calculations were performed using  a charge self-consistent LDA+DMFT approach~\cite{GKotliar:2006,KHaule:2010}  based on a full-potential linearized augmented plane wave (FP-LAPW) as implemented in the WIEN2k code.~\cite{PBlaha:2001} The generalized gradient approximation (GGA)~\cite{JPPerdew:1996} was used for the exchange-correlation functional. Hereafter, we  use interchangeably LDA and GGA as an acronym for a non-polarized generalized gradient approximation.  The spin-orbit coupling was included in a second variational way. The muffin-tin radius 2.29$a_0$ ($a_0$ being the Bohr radius), 2.03$a_0$  and 2.30$a_0$  for Fe, Ge, and Te, respectively, and a plane wave cutoff $RK_{\text{max}}=7$ were taken in calculations that included  $16\times 16 \times 3$  $\mathbf{k}$-points. Within LDA+DMFT,   to explicitly include in the DFT on-site Coulomb interactions (with strength $U$ and $J$) among Fe-3$d$ electrons, 
a clear definition of the atomic-like local orbitals is required. In this work, we used  
the weight-conserved projection procedure~\cite{KHaule:2010} to extract the local GreenÕs function for the correlated Fe  3$d$-orbitals from the full GreenÕs function defined in the DFT basis. For the DMFT,  a strong-coupling version of continuous-time quantum Monte Carlo 
(CT-QMC) method,~\cite{PWerner:2006a,PWerner:2006b,KHaule:2007} which provides numerically exact solutions, was used to solve the effective multiple-orbital quantum impurity problem self-consistently.  
 Since the DFT already includes the Hartree-term of the Coulomb interaction, we included a double-counting correction $E_{dc}=U(n_{f}^{0}-1/2) - J(n_{f}^{0}-1)/2$ with a nominal value of $n_{d}^{0}=6$ for Fe-$3d$ electrons.  For very-well defined local Fe-$3d$ orbitals,  this double-counting scheme has the virtue of numerical stability.~\cite{KHaule:2014}

\begin{figure}[t!]
\centering\includegraphics[
width=1.0\linewidth,clip]{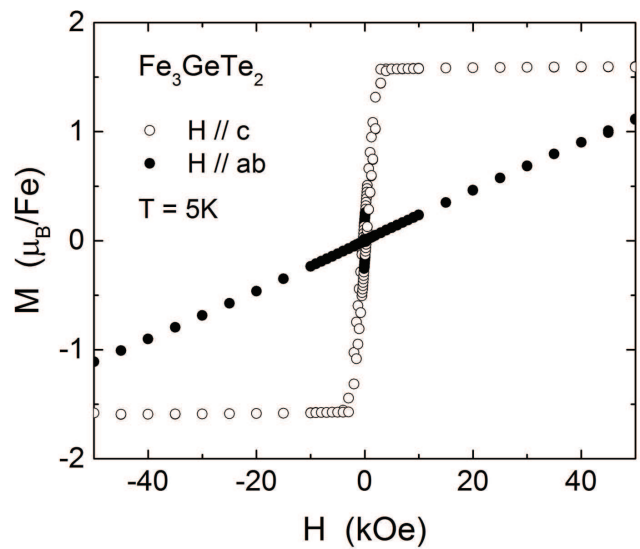}
\caption{(Color online)
Magnetization versus applied field for Fe$_3$GeTe$_2$ at 5 K with the field applied parallel and perpendicular to the $c$-axis. The magnetization is normalized by the Fe content determined by refinement of X-ray data. 
 }
\label{fig:magnetization}
\end{figure}

\section{Results and discussions}  
\label{sec:results}

Figure~\ref{fig:magnetization} plots the anisotropic magnetization at 5 K as a function of field where we see  that the $c$-axis is the easy axis, consistent with neutron diffraction results.~\cite{VVerchenko:2015} The $c$-axis saturated moment is 1.58 $\mu_B$/Fe, which is somewhat smaller than the value of 1.63 $\mu_B$/Fe reported in 
Ref.~\onlinecite{BChen:2013}, even though both samples have the same Curie temperature $T_C=223 \pm 3$ K and similar paramagnetic effective moment $\mu_{eff}=4.5$-4.7 $\mu_B$/Fe. We note that the saturated moment decreases by only about 10\% on raising the temperature to 100 K.~\cite{BChen:2013}

\begin{figure}[t!]
\centering\includegraphics[
width=1.0\linewidth,clip]{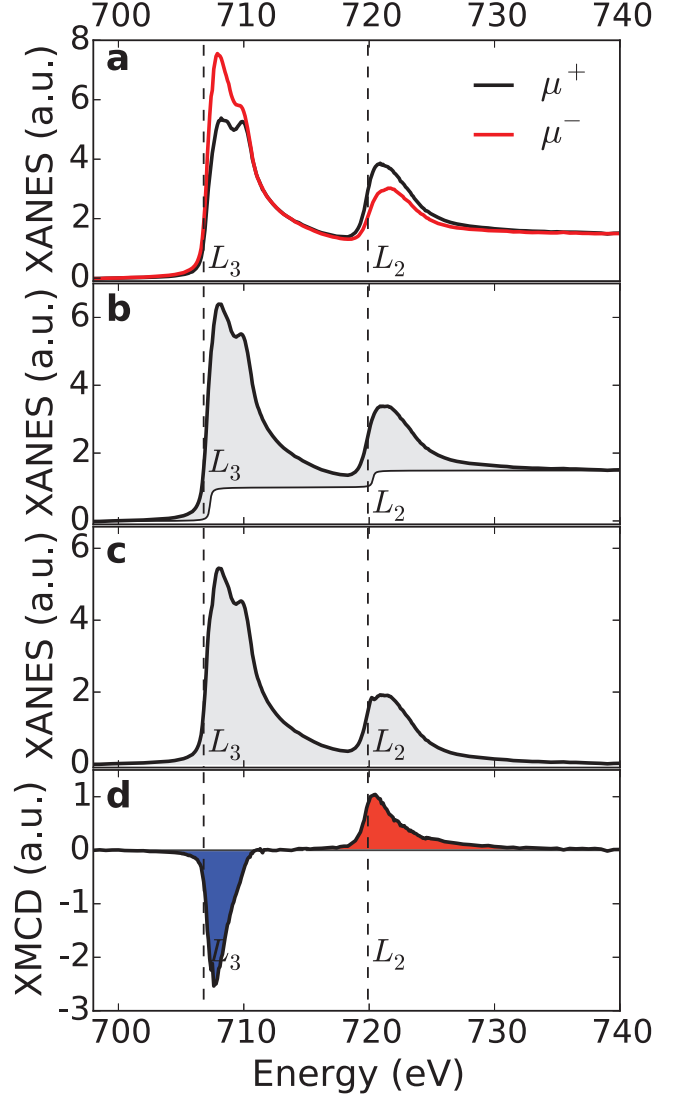}
\caption{(Color online)
X-ray absorption near edge structure (XANES) and XMCD on Fe$_3$GeTe$_2$ at a temperature $T=45$~K. (a)  Total electron yield signals on the Fe $L_3$ and $L_2$ edges in Fe$_3$GeTe$_2$ recorded with left- and right-circularly polarized x-rays, denoted with $\mu^+$ and $\mu^-$, respectively. 
(b) XANES  obtained via averaging the polarized spectra in (a). 
(c) XANES whiteline spectra at the Fe $L_3$ and $L_2$ edges.  
The grey shaded area in (b) and (c) denotes the integral $\int_{L_3+L_2}(\mu^++\mu^-)dE$ required in Eqs.~(\ref{eq:angular}) and (\ref{eq:spin}).
(d) XMCD signal $\Delta\mu=\mu^+-\mu^-$. 
 The blue and red shaded areas illustrate the integrals $\int_{L_3}\Delta\mu dE$ and $\int_{L_2}\Delta\mu dE$ over the XMCD signal at the $L_3$ and $L_2$ edges, respectively. We note that $\int_{L_3+L_2}\Delta\mu dE$  is the sum of both the blue and red shaded areas.  
}
\label{fig:xmcd}
\end{figure}

X-ray magnetic circular dichroism allows a determination of the orbital contribution. Figure~\ref{fig:xmcd}(a) shows the total electron yield signals at the Fe $L_3$ and $L_2$ edges of Fe$_3$GeTe$_2$ recorded with left- and right-circularly polarized x-rays, denoted by $\mu^+$ and $\mu^-$, respectively. The data were normalized so that the edge-step at the $L_3$ is equal to one. In Fig.~\ref{fig:xmcd}(b) we show the normalized x-ray absorption near edge structure (XANES) $\mu_0=(\mu^++\mu^-)/2$, obtained by averaging the $\mu^+$ and $\mu^-$ contributions. The background due to photo excitations into continuum states has been fitted using an \textit{ad-hoc} step-function,  where the edge-step ratio between the $L_3$ and $L_2$ edges was set to 2:1 according to the occupation of the $2p_{3/2}$ and $2p_{1/2}$ core states.~\cite{CTChen:1995} In Fig.~\ref{fig:xmcd}(c) this background has been subtracted, and only the whiteline of the recorded spectra is shown. The XMCD signal $\Delta\mu=\mu^+-\mu^-$a for Fe$_3$GeTe$_2$ is finally shown in Fig.~\ref{fig:xmcd}(d).

Using sum rules,~\cite{BTThole:1992,PCarra:1993} the angular and spin moments of Fe $\langle L_z \rangle$ and  $\langle S_z \rangle$ in the ground state of 
Fe$_3$GeTe$_2$ can be determined from the XANES and XMCD signals. Here we use the convention $\mu_L = -\langle L_ z \rangle$ and $\mu_S = -2\langle S_z\rangle$.~\cite{Gvan_der_Laan:1996} According to the sum rules the angular moment $\mu_L$ is given by 
\begin{equation}\label{eq:angular}
-\mu_L=\langle L_z \rangle = -\frac{4}{3}n_h \frac{\int_{L_3+L_2}\Delta\mu dE}{\int_{L_3+L_2}(\mu^++\mu^-)dE}\;,
\end{equation}
and
\begin{equation}\label{eq:spin}
2\langle S_z \rangle + 7\langle T_z \rangle=-n_h\frac{6\int_{L_3}\Delta\mu dE-4\int_{L_3+L_2}\Delta\mu dE}{\int_{L_3+L_2}(\mu^++\mu^-)dE} \;.
\end{equation}
Here $\mu_s = -2\langle S_z \rangle$ is the spin moment, and $\langle T_z \rangle$ denotes the magnetic dipole contribution. Further, $n_h$ is the number of holes in the 3$d$ shell, where we assume $n_h= 4$  for the 3$d^6$ configuration of Fe in Fe$_3$GeTe$_2$. The various integrals in Eqs.~(\ref{eq:angular}) and ~(\ref{eq:spin}) are denoted by the various shaded areas in Figs.~\ref{fig:xmcd}(c) and (d). See the figure caption for details.

By using Eq.~(\ref{eq:angular}), the orbital moment of Fe$_3$GeTe$_2$ is estimated to be  $\mu_L = 0.10(3)$~$\mu_B$/Fe. Because the total $3d$ iron moment is $\mu = \mu_s + \mu_L$,  we use the total moment measured by bulk magnetization measurements to deduce the spin moment. We  scale the total moment to the saturation magnetization at  $T=5$~K, $M_{s}(5\rm{~K})=1.58 \;\mu_B$/Fe,  and thus obtain the spin moment $\mu_s = 1.48(6)$~$\mu_B$/Fe. Employing Eq.~(\ref{eq:spin}), we calculate the dipole contribution to be  $\langle T_z \rangle = 0.24(5)$~$\mu_B$/Fe. For completeness, we also calculate the ratio $\mu_L/(\mu_s + 7\langle T_z\rangle) = 0.03$ using the ratio of Eqs.~(\ref{eq:angular}) and (\ref{eq:spin}). The advantage of this quantity is that it does not depend on the number of holes $n_h$ and therefore can be easily compared with other 3$d$ materials.

\begin{figure}[t!]
\centering\includegraphics[
width=1.0\linewidth,clip]{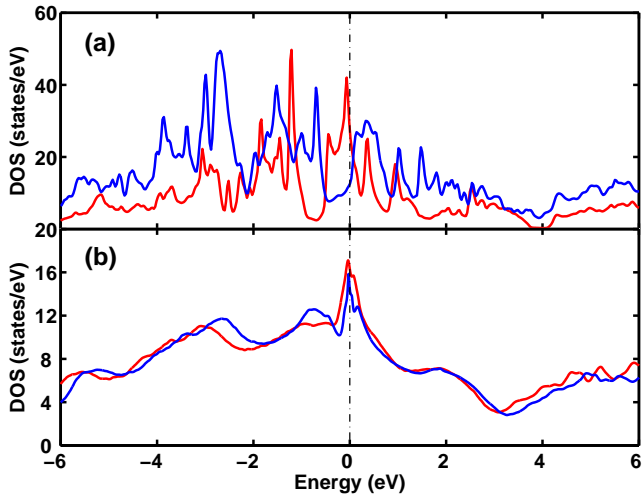}
\caption{(Color online)
Electronic density of states for Fe$_3$GeTe$_2$ in the paramagnetic (red lines) and ferromagnetic (blue lines) states calculated from the GGA (a) and GGA+DMFT (b).  The energy is measured with respect to the Fermi energy ($E=0$). In the GGA+DMFT calculations, the temperatures  $T= 232$ K for paramagnetic and $T=116$ K for ferromagnetic states were used.  We chose $T=116$ K for the FM state of Fe$_3$GeTe$_2$ in the GGA+DMFT method because it is computationally demanding. We have discussed in the text that an increase of the temperature up to 100 K only slightly reduces the moment. Therefore, the chosen temperature is indeed reasonable for the description of the FM state of Fe$_3$GeTe$_2$.
 }
\label{fig:dos}
\end{figure}

In Fig.~\ref{fig:dos}(a), we show the total density of states (DOS) as a function of energy from GGA calculations for both paramagnetic (PM) and ferromagnetic (FM) phases. In the PM phase, there is a sharp peak in the DOS (red line  in Fig.~\ref{fig:dos}(a)) appearing only about 65 meV below the Fermi energy ($E=0$). These states are dominantly of Fe-3$d$ character. The close proximity of this PM DOS peak to the Fermi energy favors an electronic instability of the PM state toward magnetic ordering, and together with the ratio of saturated to effective moment $M_s/M_{\text{eff}}$ of $\sim 0.3$, the Stoner mechanism for itinerant ferromagnetism may be applicable.  When ferromagnetic ordering is turned on in a spin-polarized GGA calculation, the DOS near the Fermi energy (blue line  in Fig.~\ref{fig:dos}(a)) is significantly suppressed but the Fe-3$d$ states still contribute dominantly, indicating the system is a ferromagnetic metal. The spin-polarized GGA calculations give a spin moment of about 2.45$\mu_B$  and 1.59$\mu_B$ and an orbital moment of about 0.08$\mu_B$  and 0.03$\mu_B$ for Fe1 and  Fe2 sites, respectively. The calculated moments agree very well with earlier electronic structure calculations.~\cite{VVerchenko:2015} The averaged total orbital moment of 0.063$\mu_B$/Fe is close to the value of 0.1$\mu_B$/Fe determined from XMCD measurements at $T=40$ K. These values are  expected from the fact that the orbital moment is proportional to the spin-orbit coupling strength in an itinerant ferromagnet.  However,  spin-polarized DFT calculations  significantly overestimate the total magnetic moment, giving an average spin moment  2.2$\mu_B$/Fe as compared to the saturated moment of  1.58 $\mu_B$/Fe measured on our crystal. The saturated moment is the sum of spin and orbital contributions, and using our value of $M_{s}$ and orbital moment from XMCD, we deduce  the spin-only  moment  of 1.48$\mu_B$/Fe.  This disagreement between theory and experiment is reminiscent of parent compounds of Fe-based superconductors, where  values of the ordered moment calculated by the DFT in the spin-density-wave phase, $M_{\text{DFT}}\approx 1.8\mu_{B}$ or higher,~\cite{DJSingh:2008} are much larger than the experimentally observed values (mostly below 1.0$\mu_B$).

\begin{figure}[t!]
\centering\includegraphics[
width=1.0\linewidth,clip]{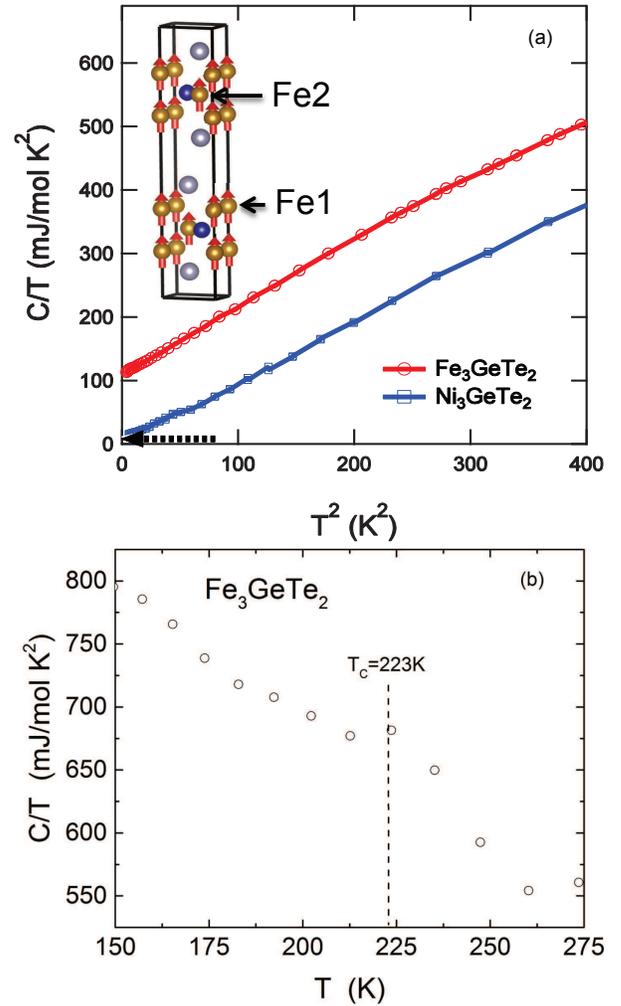}
\caption{(Color online) The formula-unit molar specific heat divided by temperatue at low (a) and high (b) temperatures. 
(a) Specific heat divided by temperature $C/T$ versus $T^2$ down to  2 K of Fe$_3$GeTe$_2$ and Ni$_3$GeTe$_2$. 
 The dashed arrow marks the value estimated from band structure calculations in the FM phase of Fe$_3$GeTe$_2$.  The inset shows the crystal structure of Fe$_3$GeTe$_2$.  Fe, Ge, and Te are represented by orange, blue, and gray spheres, respectively. 
 Each unit cell contains two Fe$_3$GeTe$_2$ layers, which alternate along the $c$-axis, and four Fe1 atoms occupy 4e sites and nominally two Fe2 atoms occupy 2c sites. (b) Specific heat divided by temperature around the Curie temperature $T_C =223 \pm 3$ K.
}
\label{fig:COT}
\end{figure}

The specific heat, plotted as $C/T$, of Fe$_3$GeTe$_2$ and isostructural Ni$_3$GeTe$_2$ is shown in Fig. \ref{fig:COT}(a). A peak  in $C/T$ of Fe$_3$GeTe$_2$  is observed at the ferromagnetic ordering temperature $T_C$= 223 K (Fig. \ref{fig:COT}(b)), but Ni$_3$GeTe$_2$ remains paramagnetic down to 2 K. The electronic contribution to the specific heat, obtained from a linear fit to \emph{C/T} versus  $T^{2}$ between 2.0 K $\leq$ T $\leq$ 11.7 K for Fe$_3$GeTe$_2$, gives a Sommerfeld coefficient 
$\gamma= 110\; \text{mJ}/\text{mol} \; \text{K}^{2}$, in agreement with previous measurements.~\cite{BChen:2013} In contrast, a linear fit
 of  $C/T$  for Ni$_3$GeTe$_2$ between 2.4 K $\leq$ T $\leq$ 11.2 K yields $\gamma= 9 \;\text{mJ}/\text{mol} \;  \text{K}^{2}$, a factor of 10 times smaller than Fe$_3$GeTe$_2$. The enhanced electronic specific heat of Fe$_3$GeTe$_2$ relative to that of Ni$_3$GeTe$_2$ suggests the presence of significant electronic correlations that may enhance the magnetic anisotropy energy. In addition, our spin-polarized GGA calculations on 
Fe$_3$GeTe$_2$ give a density of states at the Fermi energy $N(E_F)=3.5\; \text{states}/\text{eV} \cdot \text{f.u.}$, corresponding to a bare Sommerfeld coefficient $\gamma_b=\pi^2 k_B^2 N(E_F) / 3 = 8.3\; \text{mJ}/\text{mol} \; \text{K}^2$ that is very close to the measured and calculated Sommerfeld coefficient of non-magnetic Ni$_3$GeTe$_2$. This result implies an effective mass renormalization of $m^*/m_b = \gamma/\gamma_b =13.3$.
 The $T^2$ contribution to $C/T$ gives a phonon specific heat coefficient $\beta$  of $1.06\; (0.81) \;\text{mJ}/\text{mol} \; \text{K}^{4}$ for Fe$_3$GeTe$_2$ (Ni$_3$GeTe$_2$), corresponding to a Debye temperature of 222 K (243 K) [Here we use the relation $\theta_{D}$ =$\sqrt[3]{12\pi^{4}rR/(5\beta)}$ (\emph{r} is number of atoms in the formula unit and \emph{R} is the universal gas constant)].  The almost perfect fit of $C/T$ to the form of $\gamma + \beta T^2$ and the nearly identical phonon contributions of Fe$_3$GeTe$_2$ and Ni$_3$GeTe$_2$ imply a negligible magnetic contribution to the specific heat of Fe$_3$GeTe$_2$ below 20 K. This result  is expected for a ferromagnet such as Fe$_3$GeTe$_2$, where the single ion anisotropy should produce a gap in the magnon spectrum.  This comparison of theoretical and experimental values of the Sommerfeld coefficient leads to the conclusion that the quasiparticle mass is significantly enhanced and Fe$_3$GeTe$_2$ is a strongly correlated ferromagnetic metal.

\begin{figure}[t!]
\centering
\includegraphics[
width=1.0\linewidth,clip]{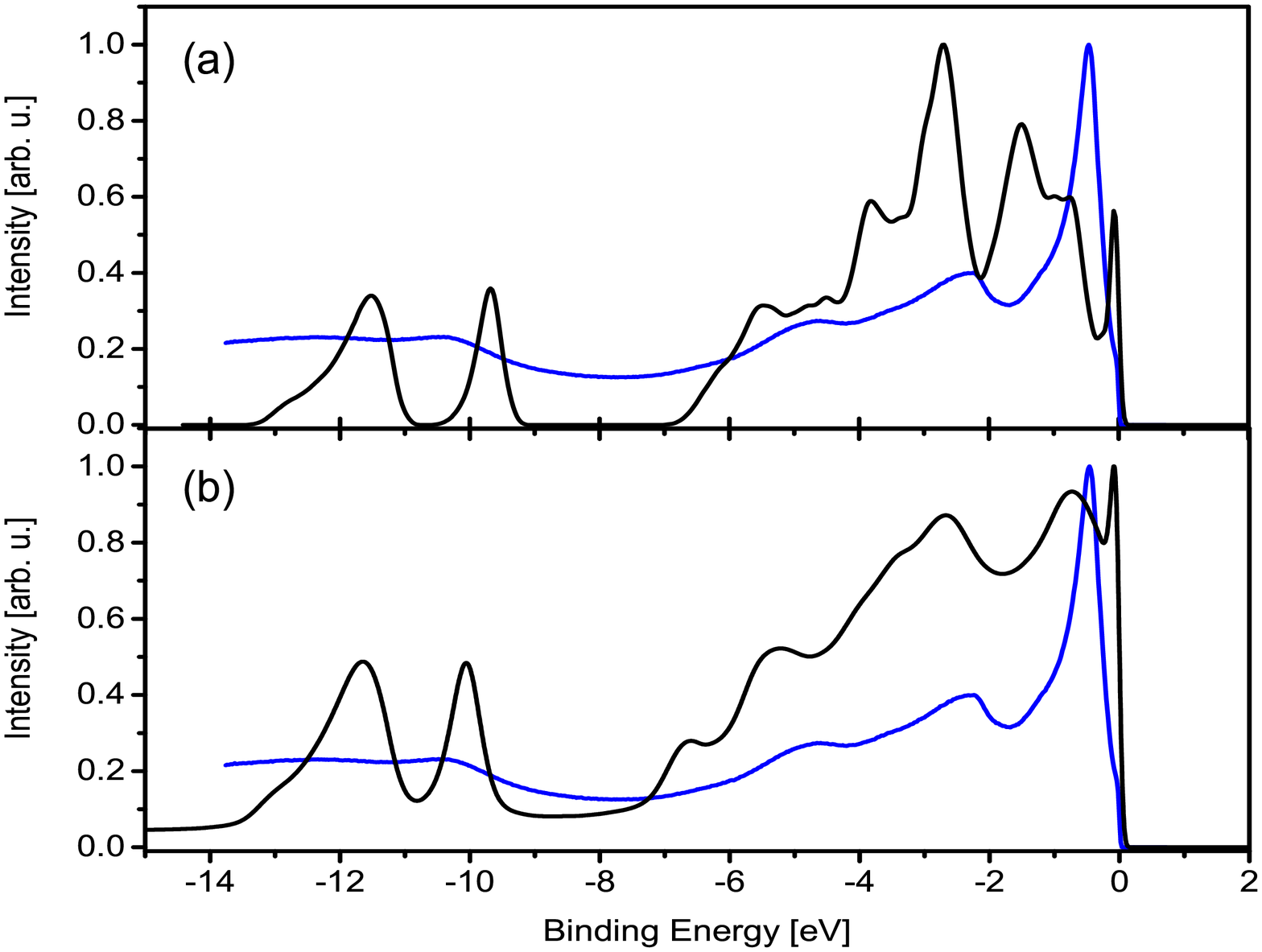}
\caption{(Color online) The spectral intensity obtained from the photoemission spectroscopy in the ferromagnetic phase of Fe$_3$GeTe$_2$ at $T=15$K.
The measured photoemission intensity (blue line) is plotted in comparison with the density of states (black lines)  obtained from spin-polarized GGA (a) and GGA+DMFT at $T=116$ K  (b) band structure  calculations.  Calculated density of states was broadened by experimental resolution and convoluted with a Fermi function at the measurement temperature. 
}
\label{fig:pes}
\end{figure}

We turn now to electronic structure calculations within LDA+DMFT, which explicitly includes  quantum temporal fluctuations. A cubic harmonic basis was used for CT-QMC simulations in the DMFT. That is,  for the PM phase, we used five distinct diagonal matrix elements in the self energy  for each of the nonequivalent  Fe; while for the FM phase, the spin degeneracy was also lifted. Since the spin-orbit coupling is weak in  Fe$_3$GeTe$_2$, this effect on the dynamical self-energy is negligible and we instead focus on the electronic correlation effects, which are dominant. 
In view of the fact that  the spin moment obtained from LDA calculations on Fe1 is much larger than that on Fe2, we chose the Hubbard interactions $U_{\text{Fe1}}=5.5$ eV and $U_{\text{Fe2}}=5$ eV and an averaged Hund's rule interaction $J=0.79$ eV in the DMFT calculations, unless specified explicitly otherwise.
These values of Coulomb interactions are close to those used in studies of iron-pnictides and chacogenides~\cite{ZPYin:2011} except that only one non-equivalent type of Fe atoms exists in their crystal structure. In addition, we chose different values of $U$ for Fe1 and Fe2 to ensure that the obtained magnetic moment on Fe1 is larger than that on Fe2 (see below for more details), a trend consistent with the spin-polarized GGA calculations. 
 Figure~\ref{fig:dos}(b) shows the total density of states for the PM phase (red line) and for the FM phase (blue line). 
Comparing results for the PM phase with those obtained from GGA calculations (Fig.~\ref{fig:dos}(a)), one can see that the DFT band within the range of about $[-2 \text{eV}, \;2 \text{eV}]$ is significantly renormalized due to the electronic correlation effects and that  the band near the Fermi energy is narrowed in  the range of $[-1\text{eV},1\text{eV}]$. Upon ferromagnetic order (blue line), the narrow band near the Fermi energy spreads to enhance intensity at energies around -0.82 eV.  However,  peak intensity near the Fermi energy does not decrease much, which differs dramatically from the results of spin-polarized GGA calculations.  

In Fig.~\ref{fig:pes}, we plot the spectral density obtained from photoemission spectroscopy on single crystal Fe$_3$GeTe$_2$ in the FM phase and  compare it to band structure calculations. Spectral features measured above and around a binding energy of 2 eV are in qualitative agreement with both spin-polarized GGA and LDA+DMFT descriptions for the FM phase. The PES also identifies a sharp peak at a binding energy of about 0.5 eV.  This is in disagreement with the spectral density obtained from spin-polarized GGA, which shows a strongly depressed intensity in the energy region between $-0.5$ eV and the Fermi level.  However,  the GGA+DMFT calculations capture a spectral intensity comparable to the PES peak, rendering a much better overall agreement with PES than the spin-polarized GGA results.

Even more significantly, calculations within the LDA+DMFT approach for the FM phase at $T=116$ K give spin moments of 1.60 $\mu_B$ for Fe1 and 1.54 $\mu_B$ for Fe2 sites. The averaged spin moment of 1.58 $\mu_B$/Fe agrees very well with values of total magnetic moment obtained from the magnetization measurement ($1.58 \mu_B$/Fe) at 5 K and with the spin-only moment (1.48 $\mu_B$/Fe) deduced from XMCD.  
This agreement between experiment and theory still is quite good when we consider that the quoted experimental moments decrease by only about 10\% at temperature of the calculations.
We have also checked the $U$-dependence of the spin moments at $T=116$ K and found they have the values of 0.31 $\mu_B$ (0.26 $\mu_B$), 0.45 $\mu_B$ (0.31 $\mu_B$), 1.32 $\mu_B$ (1.56 $\mu_B$)  for Fe1 (Fe2) when $U$ for both Fe1 and Fe2 takes the value of 3 eV, 4 eV, and 5 eV.
LDA+DMFT also allows an estimate of the effective mass enhancement due to correlation-induced renormalization.
 It is proportional to the ratio of the quasiparticle density of states to band DOS at the Fermi energy: $m^*/m_b=\tilde{\rho}(E_F)/\rho_b(E_F)$. Here $\rho_b(E_F)= \sum_{i,\alpha} w_i \rho_{b,\alpha}(E_F)$ and $\tilde{\rho}(E_F)=\sum_{i,\alpha} w_i \rho_{b,\alpha}(E_F)/z_{i,\alpha}$, where $\rho_{b,i,\alpha}$ is the partial density of states at the Fermi energy from the 10 spin orbitals for the $i$-th type of Fe atom,  $w_i$ is the number of equivalent atoms of a given type, and the quasiparticle weight  is given by $z_{i,\alpha}=[1-\partial\text{Im}\Sigma_{\alpha,i}(i\omega_n)/\partial \omega_n|_{\omega_n \rightarrow 0}]^{-1}$ with the self-energy $\Sigma_{\alpha,i}$ defined on the Matsubara frequency  $\omega_n$ axis. Our CT-QMC simulations give an estimated  effective mass of 2.61$m_{b}$.  Although this enhancement does not fully account for the large Sommerfeld coefficient,
 the theoretical value  of $m^*/m_b$ at $T=116$ K gives a lower bound for the consequences of electronic correlations in Fe$_3$GeTe$_2$.

\section{Concluding remarks}  
\label{sec:summary}
We have studied the electronic and magnetic properties of the layered  itinerant ferromagnet  Fe$_3$GeTe$_2$ through  band structure calculations within  LDA and LDA+DMFT approaches and a suite of experimental measurements including specific heat, XMCD,  and photoemission spectroscopy. We have shown that the incorporation of  quantum temporal fluctuations within LDA+DMFT gives the magnetic moments, mass enhancement, and spectral density in better agreement with  experiments than LDA. These results have demonstrated clearly that Fe$_3$GeTe$_2$ is a strongly correlated ferromagnetic metal.

\begin{acknowledgments}
We are grateful to C. D. Batista and Zhi-Ping Yin for helpful discussions. Work at LANL was performed under the auspices of the U.S.\ DOE contract No.~DE-AC52-06NA25396 through the Los Alamos LDRD program. 
Y.S.  was supported by the Wenner-Gren Foundation, and M.M. was supported by Marie Sklodowska Curie Action, International Career
Grant through the European Commission and Swedish Research Council (VR), Grant No. INCA-2014-6426.
Part of the theoretical calculations were carried out on a Linux cluster in the Center for Integrated Nanotechnologies, a DOE Office of Basic Energy Sciences user facility. 
\end{acknowledgments}

\end{document}